\documentclass[aps,prl,twocolumn,superscriptaddress]{revtex4-2}

\usepackage{bm}
\usepackage{comment}
\usepackage{amsmath,amsthm,amssymb,amsfonts}
\usepackage{mathtools}
\usepackage{bm}
\usepackage{graphicx}
\usepackage{color}
\usepackage[normalem]{ulem}
\usepackage{sidecap}
\sidecaptionvpos{figure}{t}

\usepackage{wrapfig}

\definecolor{szlcolor}{rgb}{0.8, 0, 0.6}

\newcommand{\newparagraph}[1]{\vskip0.5cm \noindent \textbf{#1}}

\newcommand{\newsubparagraph}[1]{\vskip0.25cm \noindent \textit{#1}}

\newcommand{\methodparagraph}[1]{\vskip0.25cm \noindent \textit{#1}}

\usepackage[english]{babel}

\begin{document}

\title{Noninvasive rheological inference from stable flows in confined tissues}

\author{Marc Karnat}
\affiliation{Aix Marseille Univ, Université de Toulon, CNRS, CPT (UMR 7332), Turing Centre for Living Systems, Marseille, France}

\author{Gautham Hari Narayana}
\affiliation{Université Paris Cité, CNRS, Institut Jacques Monod, F-75013 Paris, France.}

\author{Sudheer Kumar Peneti}
\affiliation{Université Paris Cité, CNRS, Institut Jacques Monod, F-75013 Paris, France.}

\author{Victoria Guglielmotti}
\affiliation{Université Paris Cité, CNRS, Institut Jacques Monod, F-75013 Paris, France.}
\affiliation{Department of Physics, Friedrich-Alexander Universität Erlangen-Nürnberg, Erlangen, Germany Max-Planck-Zentrum für Physik und Medizin, Erlangen, Germany}

\author{Qazi Saaheelur Rahaman }
\affiliation{Aix Marseille Univ, Université de Toulon, CNRS, CPT (UMR 7332), Turing Centre for Living Systems, Marseille, France}

\author{Shreyansh Jain}
\affiliation{Université Paris Cité, CNRS, Institut Jacques Monod, F-75013 Paris, France.}

\author{Benoît Ladoux}
\affiliation{Université Paris Cité, CNRS, Institut Jacques Monod, F-75013 Paris, France.}
\affiliation{Department of Physics, Friedrich-Alexander Universität Erlangen-Nürnberg, Erlangen, Germany Max-Planck-Zentrum für Physik und Medizin, Erlangen, Germany}

\author{Shao-Zhen Lin}
\affiliation{Guangdong Provincial Key Laboratory of Magnetoelectric Physics and Devices, School of Physics, Sun Yat-sen University, Guangzhou, China.}
\affiliation{Centre for Physical Mechanics and Biophysics, School of Physics, Sun Yat-sen University, Guangzhou, China.}

\author{Sham Tlili}
\affiliation{Aix Marseille Univ, IBDM (UMR 7288), Turing Centre for Living Systems, Marseille, France}

\author{René-Marc Mège}
\email{rene-marc.mege@ijm.fr}
\affiliation{Université Paris Cité, CNRS, Institut Jacques Monod, F-75013 Paris, France.}

\author{Jean-François Rupprecht}
\email{jean-francois.rupprecht@univ-amu.fr}
\affiliation{Aix Marseille Univ, Université de Toulon, CNRS, CPT (UMR 7332), Turing Centre for Living Systems, Marseille, France}
\affiliation{Aix Marseille Univ, Université de Toulon, CNRS, LAI (UMR 7333), Turing Centre for Living Systems, Marseille, France}

\date{\today}

\begin{abstract}
Quantifying the in-plane rheology of epithelial monolayers remains challenging due to the difficulty of imposing controlled shear. We introduce a self-driven, rheometer-like assay in which collective migration generates stationary shear flows, allowing rheological parameters to be inferred directly from image sequences. The assay relies on two sets of ring-shaped fibronectin patches, micropatterned in arrays for high-throughput imaging. Within isolated rings, the epithelial tissue exhibits persistent rotation, from which we infer active migration stresses and substrate friction. Within partially overlapping rings, the tissue exhibits sustained shear, from which we infer the elastic and viscous responses of the cells. The emergence of a Maxwell-like viscoelastic relation --characterized by a linear relationship between mean cell deformation and neighbor-exchange rate-- is specifically recapitulated within a wet vertex-model framework, which reproduces experimental measurements only when intercellular viscous dissipation is included alongside substrate friction. We apply our method to discriminate the respective roles of two myosin~II isoforms in tissue mechanics. Overall, by harnessing self-generated stresses instead of externally imposed ones, we propose a noninvasive route to rheological inference in migrating epithelial tissues and, more generally, in actively flowing granular materials.
\end{abstract}

\maketitle

Large-scale tissue deformations during embryonic development and tumor progression arise from the interplay between actively generated cellular stresses and the rheological properties of tissues, with cell–cell rearrangements and divisions relaxing stored deformations \cite{Lenne2022}. Tissue fluidity is increasingly recognized as being implicated in cancer growth and metastatic dissemination \cite{Lemahieu2025}, and fluidity measurements in live biopsies are proposed as diagnostic readouts \cite{Sauer2023}.

Current strategies for characterizing tissue rheology often adapt tools from materials science \cite{Petridou2019,Corominas-Murtra2021} — laser ablation to induce fractures \cite{Rauzi2010,Bosveld2012}, embedded deformable probes \cite{Mongera2018}, or externally imposed stresses \cite{Harris2012,Sadeghipour,Khalilgharibi2019,Duque2024}. However, these approaches are invasive and can be difficult to implement broadly  \cite{Petridou2019,Corominas-Murtra2021}. In contrast, image-based proxies such as the cell-shape index \(s=P/A^{1/2}\)—a dimensionless ratio of perimeter to area \cite{Bi2015}—are noninvasive.  This metric is motivated by results obtained using the vertex-model framework, predicting that increasing $s$ correlates with higher tissue fluidity, with the tissue shear modulus eventually vanishing under quasi-static shears when $s$ exceeds a critical value $s^{\star} \approx 3.8$ for isotropic, confluent, 2D simulations \cite{Bi2015,Damavandi2025}.
However, taken alone, the cell-shape index $s$ has been shown to be insufficient in several contexts (e.g. with activity or anisotropic deformations) limiting comparisons between systems \cite{Yan2019,Wang2020,Brauns2024,Damavandi2025}.

An alternative noninvasive method is provided by \emph{tissue kinematics}, a framework which has the potential to unravel rheological relations -- hence metrics of fluidity -- from image sequences alone \cite{Graner2008,Blanchard2009,Tlili2020,Lardy2025}. The macroscopic strain-rate is expressed in contributions from cell deformation and cell–cell rearrangements, $\dot{\bm{\varepsilon}} = \dot{\bm{\varepsilon}}_{r}+\dot{\bm{\varepsilon}}_{cell}$ \cite{Graner2008}. In 
epithelial monolayers, the deviatoric components of the cell deformation and cell–cell rearrangement tensors are related through a linear relation
\begin{equation}
\dot{\bm{\varepsilon}}^{\mathrm{dev}}_{r} \;=\; \frac{\bm{\varepsilon}^{\mathrm{dev}}_{\mathrm{cell}}}{\tau},
\label{eq:FittingRelaxationTime}
\end{equation}
which defines a characteristic timescale $\tau$ \cite{Tlili2020}. This linearity is consistent with a Maxwell viscoelastic liquid, where $\tau$ plays the role of a viscoelastic relaxation time, and quantifies the tissue fluidity -- with smaller $\tau$ corresponding to a more fluid state \cite{Tlili2020}. 

Yet, we lack a fundamental understanding of what sets the viscoelastic time $\tau$, and, more generally, of how kinematic relations such as Eq. (\ref{eq:FittingRelaxationTime}) emerge from cell-based dynamical models.
Here, this connection is established using a vertex model — the same computational framework that originally motivated the definition of the cell-shape index $s$. Furthermore, we compare our results with an experimental system that is designed to measure the kinematics tensor at high throughput. We grew Madin-Darby Canine Kidney (MDCK) monolayers on pairs of adjacent fibronectin rings (hereafter referred to as \textit{double-rings}). At confluency, these exhibited persistent shear (Fig.~\ref{fig:1}A), with either co- or counter-rotating flows emerging spontaneously, without a preferred chirality in either ring. The resulting cell shape and velocity fields then closely resembled those observed in a rheometer operating at a fixed strain rate. In addition, our setup provides dozens of micropatterns per experiment, while ensuring a well-controlled steady state; in contrast, the large monolayers with free edges used in Ref.~\cite{Tlili2020} involved a velocity field that depended on the local density. In addition, we employ deep-learning-based segmentation to extract the cell rearrangement tensor directly, and to validate the indirect inference method used in Ref.~\cite{Tlili2020}.

We then interpreted the emergence of a tissue kinematics relation Eq. (\ref{eq:FittingRelaxationTime}) within a dynamical model. We use an active viscous vertex-model framework in which cell motion results from active migration, cell deformability, and dissipation through both cell–cell interfaces and the substrate. Here, we exploit another advantage of the ring geometry setup, which is to enable a two-step calibration of parameters. First, we calibrated the active migration forces and substrate friction using data of migrating tissues on isolated rings. Second, we calibrated the elastic and viscous components associated with cell deformability, using  our double-ring setup. We find that a \emph{dry}, substrate-friction–based model fails to recapitulate the data, whereas a \emph{wet} model (based on intercellular viscous dissipation) does, identifying junctional viscosity as a necessary dissipation channel. Our model further predicts that the viscoelastic time $\tau$ introduced in Eq. (\ref{eq:FittingRelaxationTime}) decreases with increasing active traction; in agreement, we observe a larger $\tau$ when nonmuscle myosin II activity is impaired. Moreover, our analysis suggests distinct roles for myosin isoforms: myosin IIA sets the active traction magnitude, while myosin IIB also enhances the tissue yield stress. In the Perspectives section, we discuss how the image-based kinematics and mechanistic modeling presented here could be more generally applied to flowing, active granular materials.


\section*{Results}
\vskip-0.5cm
\newparagraph{Experimental observations} 

\begin{figure}[t!]
\centering
\includegraphics[width=8.5cm]{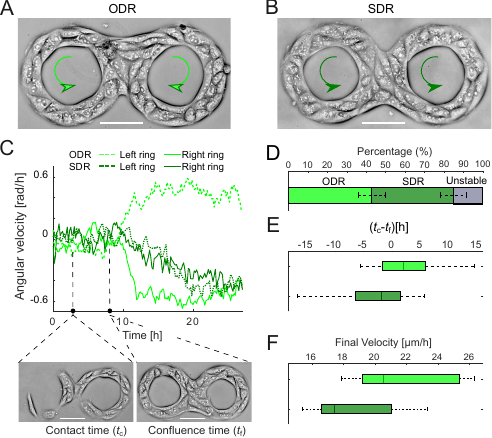}
\caption{\textbf{Spontaneous shear at double-rings overlap}. 
(A-B) Madin-Darby Canine Kidney (MDCK) cells are deposited on confining fibronectin patterns that are shaped as two rings in close contact. Upon reaching confluency, the directions of rotations (DR) within each ring are either
(A) opposite (ODR) or
(B) same (SDR) patterns 
(scale bars: $100 \, \mu\mathrm{m}$.  
(C) Average angular velocity within each ring in ODR (light green) and SDR (dark green) experiments. We report the timing of the ring-ring contact time ($t_{\mathrm{c}}$) and the confluence time ($t_{\mathrm{f}}$) in the ODR experiment.
(D) Observation frequency of the ODR, SDR, and unstable rotation patterns, 
(E) Difference ($t_{\mathrm{c}}-t_{\mathrm{f}}$) between the ring-ring contact time ($t_{\mathrm{c}}$) and the confluence time ($t_{\mathrm{f}}$) in the ODR (light green) and SDR (dark green) cases. 
(F) Mean velocities at steady state in the ODR (light green) and SDR (dark green) cases (averaged over $n=23$ rings).} 
\label{fig:1}
\end{figure}

\newsubparagraph{Spontaneous shear} Fibronectin was first micropatterned into single-ring, as in \cite{Jain2020}, and then into double-ring geometries. In the double-ring geometry, Fig. \ref{fig:1}A-B, the ring edges are separated by a distance of $3\mu$m; each display an external radius of $R_{\mathrm{out}} = 100\mu$m and an internal one of $R_{\mathrm{in}} = 90\mu$m; thus the distance between the centers of the rings is $D = 203\mu$m.

Typically, at $10 \mathrm{h}$ post-plating, a collective rotation emerged within each ring. In most cases, these rotations were sustained, lasting beyond $20 \mathrm{h}$, see Fig. \ref{fig:1}C, SI Fig. S1 and \textbf{Movies 1, 2}. Depending on the signs of each rotation, we observed (1) opposite direction of rotations (ODR),  Fig. \ref{fig:1}A, or (2) same direction of rotations (SDR), Fig. \ref{fig:1}B. Among the $n=111$ double-rings observed, $47$ were ODR, $46$ were SDR, and $18$ were unstable as they did not reach a stable flow pattern, see Fig. \ref{fig:1}D; the fraction of ODR was not significantly different from that in SDR: 50.5{$\pm 5\%$} against 49.5{$\pm 5\%$}, respectively (mean $\pm$ standard deviation), among stable patterns.

We recorded the time at which stable adhesion between the two rings formed (denoted $t_c$) as well as the time at which both rings were fully covered by cells (denoted $t_f$). We find that, on average, the contact time was slightly lower than the confluence time ($t_c - t_f<0$) in the SDR case, but slightly higher ($t_c - t_f>0$) in the ODR case; see Fig. \ref{fig:1}E ($p<0.1$; $n_{\mathrm{exp}} = 23$). 

We then reported the average rotation velocity at steady state, see Fig. \ref{fig:1}F and Methods. The velocity was significantly faster in the ODR case (at $21.6\pm3.2 \ \rm \mu m.h^{-1}$) than in the SDR ($18.7\pm2.8 \ \rm \mu m.h^{-1}$); the difference is statistically significant ($p = 0.04$; $n_{\mathrm{exp}} = 23$). 


\begin{figure}[t!]
\centering
\includegraphics[width=8.5cm]{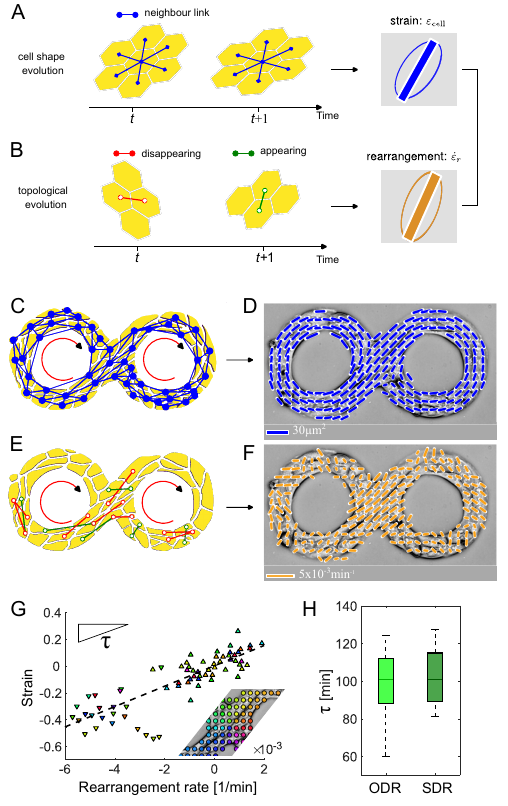}
\caption{\textbf{Cell deformability estimation} 
(A) Cell shape: connectivity graph between cell barycenters; time-averaging defines the local strain tensor $\bm{\varepsilon}_{\mathrm{cell}}$, represented as ellipses with axes proportional to eigenvalues (“coffee bean” representation \cite{Graner2008}).
(B) Topological changes: junction appearance (green) and disappearance (red); time-averaging yields the rearrangement rate tensor $\dot{\bm{\varepsilon}}_{r}$, see Eq.~(\ref{eq:Tfield}) .
(C–H) Same Direction of Rotation (SDR) experiment.
(C) Instantaneous connectivity graph.
(D) Elongation axis, $\bm{\varepsilon}^{\mathrm{dev}}_{\mathrm{cell}}$: eigenvector orientation and eigenvalue amplitude of the deviatoric tensor, see (A).
(E) Instantaneous appearance (green) and disappearance (red) axes.
(F) Rearrangement rate tensor, $\dot{\bm{\varepsilon}}^{\mathrm{dev}}_{r}$ as defined in (B).
(G) Component-to-component plot of $\bm{\varepsilon}^{\mathrm{dev}}_{\mathrm{cell}}$ vs. $\dot{\bm{\varepsilon}}^{\mathrm{dev}}_{r}$, expressed in terms of their diagonal (XX, YY, upward triangles) and off-diagonal (XY, downward) components. The viscoelastic time $\tau$ is the slope (dashed regression line). The symbol color codes for the position at which the tensors are sampled within the ring–ring contact region (inset).
(H) Distribution of viscoelastic times in ODR (light green) and SDR (dark green) modes ($n=23$ rings).} 
\label{fig:2}
\end{figure}

\newsubparagraph{Emergent Maxwell in experiments} We observed extensive cell deformations between the two rings. In the SDR mode, cells are subjected to large simple shear, while in the ODR, cells are subjected to large stretch deformations. Cell division was inhibited using Mitomycin C (see Methods) as in Refs. \cite{Jain2020,Tlili2020}, hence the persistent pattern of cell deformation observed between the two rings arises solely from the competition between cellular deformation (strain) induced by collective flow and relaxation driven by cell rearrangements. Oriented cell rearrangements typically occur in regions with large strain, as discussed in Refs.  \cite{Graner2008,Tlili2020,DUCLUT2021203746}.
To test such correlation in experiments, we estimated the averaged strain field $\bm{\varepsilon}_{\rm cell}$ 
through a custom-made segmentation pipeline, see Fig. \ref{fig:2}C-D and Methods. We then estimated the averaged rearrangement rate tensor field $\dot{\bm{\varepsilon}}_{r}$ using a cell tracking algorithm \cite{ulicna2021automated}, see Fig. \ref{fig:2}E-F and Methods. 
Following \cite{Tlili2020}, we compared the value of these tensors component by component, Fig. \ref{fig:2}G. We observed a linear relation between the deviatoric components of these two tensors, corresponding to Eq. (\ref{eq:FittingRelaxationTime}).
We find that the viscoelastic times are not significantly different ($p = 0.58$) in the ODR and SDR cases, with $\tau = 98 \pm 19 \ \rm min$ and  $102 \pm 17 \ \rm min$ (mean$\pm$std), respectively, see Fig. \ref{fig:2}H. Here, the error magnitude for $\tau$ is obtained from the distribution of fitted $\tau$ across individual double-rings. We do not take into account errors in a systematic expansion (decomposition) of one tensor into a multiple of the other.

\vskip0.5cm
\newparagraph{Wet vertex model: emergent Maxwell rheology}

\newsubparagraph{Model definition} 
To explore the relationship between the viscoelastic time $\tau$ and cellular properties, such as the ability of cells to migrate on the substrate or deform, we turned to a cell-based computational model of epithelial tissues. Specifically, we employed a vertex model that predicts the dynamics of tricellular junctions, modelled as points called vertices, in terms of cell-based forces. The force balance reads: $\bm{F}_i^{(\text{dissipation})}  + \bm{F}_i^{(\text{elastic})} + \bm{F}_i^{(\text{{active}})} = \bm{0}$, where: (1) $\bm{F}_i^{(\text{dissipation})}$ is a function of the vertices' velocities (extensively discussed in a subsequent paragraph, \cite{Fu2024}); 
(2) $\bm{F}_i^{(\text{elastic})}$ represents the elastic forces that account for the mechanical regulation of the cell shape \cite{honda1980much, fletcher2014vertex, alt2017vertex, lin2023structure, lin2018dynamic}, which are assumed to derive from the mechanical energy: $E = \sum_{J} \frac{1}{2} K_A (A_J - A_0)^2 + \sum_{J} \frac{1}{2} K_P (P_J - P_0)^2$, with $A_J$ and $P_J$ the area and the perimeter of the $J$-th cell, respectively, and $A_0$ and $P_0$ are the preferred area and preferred perimeter, respectively; (3) $\bm{F}_i^{(\text{active})} = \sum_J \bm{F}_J^{(\text{active})} $ is the sum of the active force defined within each neighbouring cell $J$, $\bm{F}_J^{(\text{active})} = F_a (\cos\theta_J,\sin\theta_J)$, where $F_a$ is a constant traction force magnitude and $\theta_J$ is the cell polarity orientation. Following Ref. \cite{lin2018dynamic}, we consider a self-alignment mechanism \cite{Vercruysse2024,RevModPhys.97.015007}, whereby the cell polarity orientation $\theta_J$ tends to align with the velocity of the neighbor cells
\begin{align}
\frac{d\theta_J}{dt} &= \frac{\mu_a }{n_J} \sum_{K\in\text{neighbor}} \sin[\theta_K^{(\text{vel})} - \theta_J],
\end{align}
where $1/\mu_a$ represents the characteristic time of alignment; $\theta_K^{(\text{vel})} = \arg(\bm{v}_K)$ refers to the argument angle of the velocity $\bm{v}_K$ of the neighbouring cell $K$   \cite{lin2018dynamic}.


\begin{SCfigure*}[\sidecaptionrelwidth][h!]
\centering
\includegraphics[width=10cm]{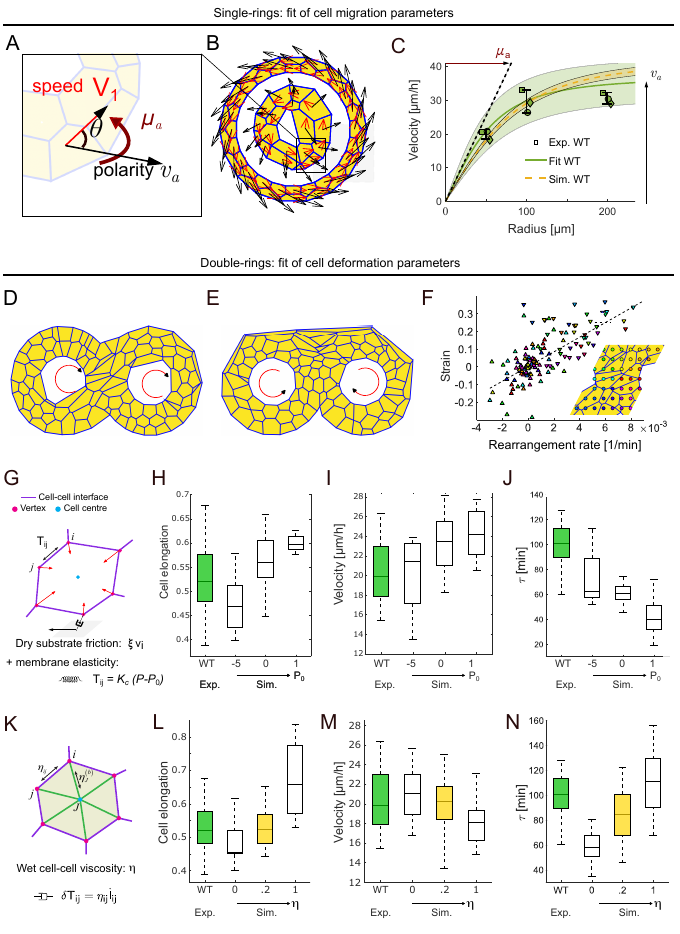}
\caption{\textbf{Cell-based simulations with self-alignment interactions and cell-cell viscosity recapitulate experiments} 
(A-C) single-ring geometry; 
(A) Sketch of the velocity (red arrow) to polarity (black arrow) coupling ($\mu_a$) and associated mismatch angle $\theta$.
(B) Simulation output: the velocity-polarity mismatch angle $\theta$ is larger on the smaller ring, with the 2 non-interacting concentric rings mimicking the high-throughput setup of Ref.  \cite{Jain2020}
(C) Mean velocity as a function of the root mean square of the external and internal ring radius, $r= \sqrt{(R_{\mathrm{in}}^2+R_{\mathrm{out}}^2)/2}$, in experiments (colored symbols), analytical fit to experiment (solid line: best model; shaded areas: best model $80\%$ confidence interval), and simulations of the vertex model (dashed lines)
(D-N) double-ring geometry; simulations with
(D) same (SDR) and
(E) opposite (ODR) direction of rotation patterns.
(F) Component-to-component relationship between cell rearrangement and strain deviatoric tensors. Up triangles for the diagonal (XX,YY) components and down triangles for the off-diagonal (XY) ones; the filling color corresponds to the location within the ring-ring contact  (inset: map of the color to a location within the ring-ring contact region; orientation as in D). The viscoelastic time $\tau$ is given by the slope (dashed line).
(G) Sketch of elastic forces that resist cell deformation (red arrows); the difference between the perimeter of the cell $P$ and the target perimeter $P_0$ sets the cell level of cell junction tension;
(H-J) Simulations with only substrate friction (dry model): cell elongation (H), velocity (I), and viscoelastic time (J) in experiments (green) and simulations with increasing preferred perimeter $P_0$. 
(K) Sketch of the cell-cell viscosity $\eta$ (wet model).
(L-N) Simulations with varied cell-cell viscosity lead to a quantitative fit of cell elongation (L), velocity (M), and viscoelastic time (N) to experiments (green) and simulations; our best parameter set (see Table S1 \cite{SM}) is indicated in yellow; $n_{\mathrm{exp}} = 23$, $n_{\mathrm{sim}} = 16$ for each set.
} 
\label{fig:3}
\end{SCfigure*}

\newsubparagraph{Single-ring confinement and calibration of active forces} The first step of our method consists in considering isolated rings, which allows us to fit the cell migration parameters to experiments. 

Confining vertices to a single-ring (see Methods), our simulations display a persistent rotation.
As the ring radius is decreased, we observe an increased mismatch between the velocity and polarity directions; see Fig. \ref{fig:3}A-B. This results in a rotation speed that is an increasing function of the ring radius, as previously observed in experiments on WT MDCK \cite{Jain2020}, see Fig. \ref{fig:3}C. We developed an analytical model for the evolution of the velocity-to-polarity mismatch angle (see details in the Supplemental Material \cite{SM}). We find that, at steady state, the resulting rotation velocity $V_1$ within a thin radius $r$ reads
\begin{equation} \label{eq:vfinal}
V_1(r) = v_a \frac{\mu_a r}{\sqrt{\mu_a^2 r^2 + v_a^2}},
\end{equation}
where $v_a = F_a/\xi$ is the maximum achievable velocity, attained in the large ring size $r \gg r_c = v_a/\mu_a$, limit. 
In contrast, in the small ring size limit $r \ll r_c$ the steady-state velocity scales as $\sim \mu_a r$, because the velocity direction changes too fast for the polar angle to follow. 

The velocity in vertex-model simulations match Eq. (\ref{eq:vfinal}) analytical prediction within 5\%, for a large set of $v_a$ and $\mu_a$ values (see Supplemental Material \cite{SM}). Experiments also fit to Eq. (\ref{eq:vfinal}), with a fit that yields the values: $v_a = 32.9 \pm 4.0 \ \rm \mu m.h^{-1}$, $\mu_a= 0.265 \pm 0.078$ h$^{-1}$ ($n_{\mathrm{exp}} = 9$) in the WT experiments. We used these values to calibrate our WT case parameter set, see Table S1 (Supplemental Material \cite{SM}).

\newsubparagraph{Double-ring confinement and estimation of elastic and viscous forces} The second step of our method consists in considering double-rings, which allow us to calibrate the cell deformability forces. 

We first considered the dry vertex model in which dissipation, $\bm{F}_i^{(\text{dissipation})}$,  is restricted to friction to the substrate $\bm{F}_i^{(\text{friction})} = -\xi \bm{v}_i$. Using the parameters deduced from the single-ring experiment, we find that the ODR and SDR modes of rotation spontaneously emerged, with nearly equal proportions as in experiments. However, within the dry vertex model framework, we were unable to find quantitative agreement with other experimental measures, such as cell elongation, velocity, and viscoelastic time; see Fig. \ref{fig:3}H-J. 

We reached a quantitative agreement with experiments upon considering a wet vertex model, i.e., that includes viscous dissipation in the form introduced in Ref. \cite{Fu2024}: $\bm{F}_i^{(\text{viscous})} = \eta \sum_{j} \bm{t}_{(i,j)} \cdot (\bm{v}_j - \bm{v}_i) \bm{t}_{(i,j)}$ where $\bm{t}_{(i,j)}$  is the unit vector from vertex $i$ to vertex $j$, indicating the direction between the vertices $i$ and $j$ and $\eta$ represents the viscous modulus along the cell-cell junction (see Supplemental Material \cite{SM}).

With our best simulation parameter set, we obtain: (1) almost equilibrated ODR and SDR populations, with 18 ODR against 15 SDR out of 33 simulations, corresponding to fractions at 54.5{$\pm 8\%$} versus 45.5{$\pm 8\%$}, respectively (mean$\pm$standard deviation), (2) velocity values quantitatively consistent with experiments, with significantly higher values in the ODR ($20.9 \pm 2.9 \ \rm \mu m.h^{-1}$) than in the SDR ($18.6 \pm 2.0 \ \rm \mu m.h^{-1}$) mode ($p = 0.04$, $n_{\mathrm{sim}} = 33$), (3) comparable viscoelastic times as in experiments, with $\tau = 84.1  \pm  17.7 \ \rm min$ for ODR and $\tau = 81.6  \pm 24.5 \ \rm min$ for SDR; as in experiments, these values were not significantly different ($p = 0.77$, $n_{\mathrm{sim}} = 33$). Representative simulations are presented in \textbf{Movie 3}.


\newparagraph{Application to myosin II perturbations} We applied our method to MDCK cell experiments in which the activity of non-muscle myosin IIA (NMIIA) and myosin IIB (NMIIB) molecular motors were reduced using shRNA techniques, called ShIIA and ShIIB, respectively
\cite{10.7554/eLife.46599}.  

\begin{figure}[t!]
\centering
\includegraphics[width=7cm]{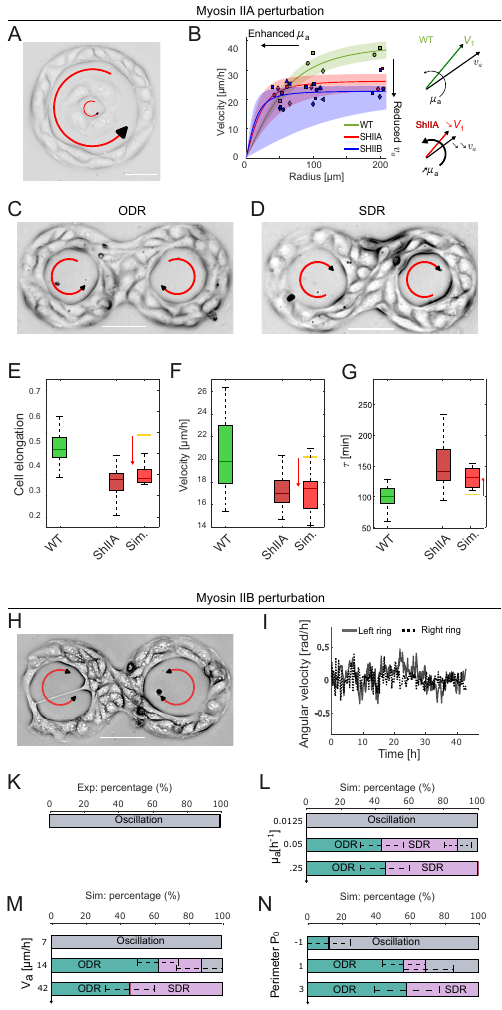}
\caption{\textbf{Myosin perturbations}. 
(A–G) Myosin-IIA silencing (ShIIA).
(A–B) Single-ring.
(A) Brightfield image: concentric rings with increasing radii migrate at increasing speeds (scale bar: $50 \, \mu\mathrm{m}$).
(B) Left: Angular velocity vs. the rms ring radius $r= \sqrt{(R_{\mathrm{in}}^2+R_{\mathrm{out}}^2)/2}$ (red, $n=3$) fitted with Eq. (\ref{eq:vfinal}); WT (green). Right: Model sketch with $V_1$ (ring velocity), $v_a$ (active traction), and $1/\mu_a$ (active coupling time) reduced in ShIIA and ShIIB (red and blue curves, respectively) as compared to WT (green) (shaded areas: model $80\%$ confidence interval).
(C–G) Double-rings.
(C–D) Brightfield image of ShIIA-treated tissues displaying (C) stable opposite (ODR) and (D) same (SDR) direction of rotation modes (scale bar: $100 \, \mu\mathrm{m}$).
(E–G) Statistics of elongation, velocity, and viscoelastic time: experiments (WT, dark green; ShIIA, dark red) and simulations (light red: best ShIIA fit). Red arrows highlight deviations from WT simulations.
(B; H–N) Myosin-IIB silencing (ShIIB).
(H) Brightfield image (scale bar: $100 \, \mu\mathrm{m}$).
(I) Angular velocity, showing no net sign.
(K–M) The oscillation phase dominates in (K) experiments and (L–N) simulations with (L) weaker alignment $\mu_a$, (M) lower traction $v_a$, or (N) smaller preferred perimeter $P_0$ (higher junctional tension) than in the control simulations (Table S1, \cite{SM}). 
} 
\label{fig:4}
\end{figure}

\newsubparagraph{Myosin IIA} Within single-rings, NMIIA-silenced cells (ShIIA) migrated persistently, see \textbf{Movie 4} at a lower velocity than  the WT; the velocity plateau is also reached at a lower critical radius, see Fig. \ref{fig:4}A. The fit to Eq. (\ref{eq:vfinal}) yields $v_a = 26.3 \pm 2 \ \rm  \mu m.h^{-1}$ and $\mu_a = 0.8 \pm 0.1 \ \rm h^{-1}$ in ShIIA case ($n_{\mathrm{exp,ShIIA}} = 3$; against $v_a = 39.2 \pm 3.0 \ \rm  \mu m.h^{-1}$ and $\mu_a= 0.5 \pm 0.05 \ \rm h^{-1}$ in the WT case).

Within double-rings, ShIIA cells displayed both the SDR and ODR migration modes (\textbf{Movies 5, 6}), with $67$ SDR, $50$ ODR and $2$ unstable. The mean velocity, at $17.1 \pm 1.2 \ \rm \mu m.h^{-1}$, was significantly reduced compared to the WT case ($p = 6.6 \times 10^{-8}$, with $n_{\mathrm{exp}, \mathrm{ShIIA}} = 46$, $n_{\mathrm{exp}, \mathrm{WT}} = 23$, combining data from ODR and SDR sets).
The viscoelastic time of ShIIA cells increased to $149\pm38 \ \rm min$, which is significantly higher than that of WT ($p =1.6 \times 10^{-7}$). This increase is consistent with simulations whose active traction ($v_a$) and alignment time ($1/\mu_a$) correspond to single-ring fit data, while keeping all other parameters fixed as in the WT simulation set, see Fig. \ref{fig:4} and \textbf{Movie 7}. The velocity of the rings decreased from a value of $19.8 \pm 2.8 \ \rm \mu m.h^{-1}$ to a value of $16.7 \pm 2.9 \ \rm \mu m.h^{-1}$ (combining ODR and SDR, $p = 8.1 \times 10^{-3}$); $n_{\mathrm{sim}} = 24$ simulations for each set of parameters). The viscoelastic time changed from $83 \pm 20 \ \rm min$ to $135 \pm 28 \ \rm min$ with lower activity ($p = 1.8 \times 10^{-6}$). Therefore, our model suggests that impaired motility alone, without any significant modulation of cell deformability, is sufficient to explain the difference in the ShIIA case compared to the WT one. 


\newsubparagraph{Myosin IIB} 
Within single-rings, NMIIB-silenced cells (ShIIB) undergo sustained rotations with velocities reduced compared to the WT case, see Fig. \ref{fig:4}A. A fit of the velocity-radius relation Eq. (\ref{eq:vfinal}) yields similar $v_a$ and $\mu_a$ values as in the ShIIA case, with $v_a = 22.7 \pm 2 \, \mu\mathrm{m}.\mathrm{h}^{-1}$ and $\mu_a = 0.9 \pm 0.3  \ \rm h^{-1}$.

However, in double-rings, ShIIB-tissues do not sustain persistent rotations, but rather rapid oscillations, see Fig. \ref{fig:4}H-K and \textbf{Movie 8}. Similar oscillations emerge in simulations upon decreasing either the active traction ($v_a$), cell deformability ($P_0$) or polarity to align with velocity ($1/\mu_a$), see Fig. \ref{fig:4}M-N and \textbf{Movie 9}. Indeed, the passive response of tissue is that of a yield stress material, with a yield stress $\sigma_Y$ that decreases linearly with the target perimeter $p_0 = P_0 / \sqrt{A_0}$, and vanishes at a critical value $p^{\ast}_0 \sim 4$, as previously reported \cite{lin2023structure}.  Since the $v_a$ and $\mu_a$ seem unchanged as compared to the ShIIA case, the failure to reach a stable shear pattern and the emergence of oscillations indicate a decrease in cell deformability in the ShIIB case.

\section*{Discussion}
\vskip-0.5cm
\newparagraph{Discussion: model parameters screening}

\newsubparagraph{Viscoelastic time scaling with the cell-cell tension} We observe that the viscoelastic time $\tau$ decreases (resp. increases) with the target perimeter $P_0$ (resp. junctional tension), see Fig. \ref{fig:3}H-I. This is consistent with tension increasing the height of energy barriers to rearrangements \cite{Marmottant2009,Bi2015,Bi2016,Hertaeg2024}; yet considering a scaling in the form $\tau \propto \eta_{\mathrm{app}}/G_{\mathrm{app}}$ \cite{Bonfanti2020} is not illuminating here, as tension enters in the global viscosity $\eta_{\mathrm{app}}$  \cite{Marmottant2009} and elasticity $G_{\mathrm{app}} \propto P^{\star}_0 - P_0$ \cite{Bi2015}. The cell shape index $s = P/\sqrt{A}$ is such that lower $s$ (higher tension) corresponds to more solid-like tissues. In this sense, the mirror dependence of $1/\tau$ and $s$ on $P_0$ indicates that $1/\tau$ could be viewed as an alternative, image-sequence-based proxy of tissue fluidity.

\begin{figure}[t!]
		\centering
		\includegraphics[width=8cm]{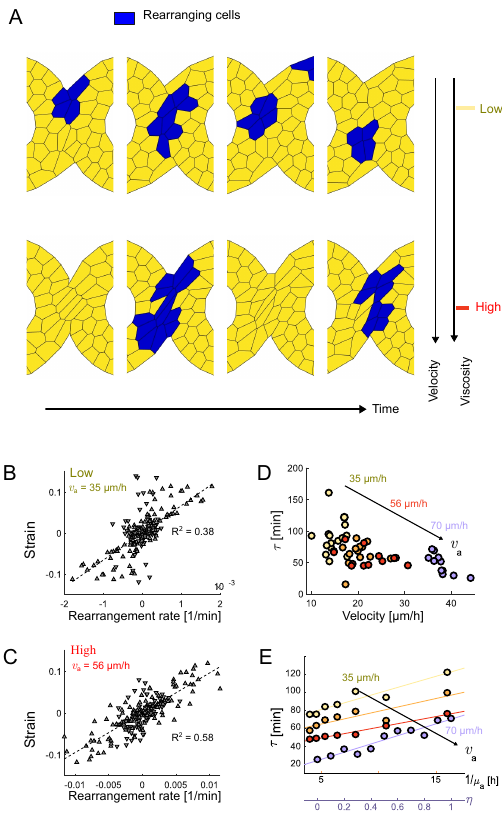}
				\caption[cell-cell junction viscosity influence on the system]{\textbf{Emergent Maxwell model interpretation.} 
    (A) Time evolution of simulations with a highlight on cells undergoing a rearrangement (colored in blue) in an SDR case with either (top) low cell-cell viscosity or low rotation speed and (bottom) high cell-cell viscosity or high rotation speeds (parameters in Supplemental Material \cite{SM}). 
    (B-C) Component-to-component relationship in the deviator of the cell rearrangement rate ($\dot{\varepsilon}^{\mathrm{dev}}_{r}$) and strain ($\varepsilon^{\mathrm{dev}}_{\mathrm{cell}}$) tensors, in (B) the low and (C) the high-velocity cases. Triangles oriented upward for the diagonal components (XX, YY) and downward for the off-diagonal ones (XY). 
    (D) Viscoelastic times for simulations with increased active traction forces, $v_a$; each dot represents the mean over simulations at a fixed polar-velocity coupling $1/\mu_a$ value.
    (E) Viscoelastic time as a function of (axis 1, yellow to red lines) the polar-velocity coupling $1/\mu_a$, or (axis 2, magenta line) the cell-cell junction viscosity $\eta$.}
				\label{fig:5}
\end{figure}

\newsubparagraph{Viscoelastic time scaling with viscosity} We find that viscosity has the opposite effect to tension, as the viscoelastic time increases  with the cell-cell junction viscosity, see Fig. \ref{fig:5}E. In the absence of cell-cell viscosity ($\eta = 0$), the velocity ($18.5 \pm 2.6 \ \rm \mu m.h^{-1}$) was comparable to the velocity ($21.1 \pm 2.9\mu$m/h) measured in our best WT parameter set ($p=0.19$), 
while both the cell elongation (at $0.48 \pm 0.06$) and viscoelastic time ($58 \pm 13 \ \rm min$) were significantly reduced ($p < 0.001$). 

Our interpretation is the following: as viscous dissipation precludes fast junction reorganization, the imposed strain-rate is shared more equally between junctions than in the absence of viscous dissipation. In agreement with this interpretation, we observe that rearrangements are less stereotypic and are more spatially distributed in the absence of viscous dissipation; see Fig. \ref{fig:5}A and \textbf{Movie 10}.
In addition, we find that the quality ($R^2$) of the linear fit Eq. (\ref{eq:FittingRelaxationTime}) is an increasing function of the viscosity $\eta$, see Fig. \ref{fig:5}B-C and Fig. S9 \cite{SM}. This allows us to reconsider the failure of the dry vertex model at $\eta = 0$. Beyond failing to match the exact values of the cell elongation, velocities, and viscoelastic together, the agreement to the linear Maxwell-like relation between cell deformation and rearrangement rate is also much poorer, see the corresponding low $R^2$-values in SM \cite{SM}). Such systematic dependence of both the value of the slope and the quality of the linear fit on $\eta$ is a nontrivial prediction of our wet model.


\newsubparagraph{Viscoelastic time scaling with active traction} We then considered a sweep in the active traction parameters ($v_a \in [35, 56] \ \rm \mu m.h^{-1}$ and $\mu_a \in [0.0625,0.25] \ \rm h^{-1}$) at a fixed cell-cell junction viscosity $\eta = 0.2$ (WT case) and for a set of junction viscosities ($\eta \in [0,1]$) at  fixed active traction forces ($v_a = 70 \ \rm \mu m.h^{-1}$ and $\mu_a = 0.25 \ \rm h^{-1}$). 
We observe that the viscoelastic time decreases with velocity toward constant values at high velocities. As expected, the viscoelastic time increases linearly with the interfacial viscosity. Furthermore, increasing the characteristic alignment time $1/\mu_a$ also leads to an increase in the viscoelastic time. These numerical results indicate that, under the ShIIA condition, the increase in viscoelastic time $\tau$ should solely be attributed to the decrease in $v_a$, and not to the decrease in $1/\mu_a$.

Such shear-thinning behavior is consistent with the existence of a yield stress  \cite{Marmottant2009,Ishihara2017,Hertaeg2024,DUCLUT2021203746}. We also demonstrated such finite yield stress through the existence of a finite active traction needed to generate the double-ring shear (see \textit{Myosin IIB} paragraph in the Results section). 

Here, we briefly sketch why a finite yield stress can explain such shear-thinning behavior. The sheared tissue is described as a material of elastic modulus $G$ in parallel with a viscous damper describing local rearrangements. Each T1 event relaxes a finite elastic stress $G \, \delta \varepsilon_{T_1}$. Viscous dissipation then occurs with a characteristic stress contribution $\eta \dot{\gamma} / \delta \varepsilon_{T_1}$, where $\dot{\gamma}$ is the imposed shear rate. This leads to a stress relation $\sigma(\dot{\gamma}) \;=\; G \, \delta \varepsilon_{T_1} \;+\; (\eta/\delta \varepsilon_{T_1}) \, \dot{\gamma}$; the apparent viscosity, $\eta_{\mathrm{app}} = \sigma(\dot{\gamma})/\dot{\gamma}$ then reads
\begin{equation} \label{eq:apperentshearthinning}
\eta_{\mathrm{app}}(\dot{\gamma}) \;=\; \frac{G \, \delta \varepsilon_{T_1}}{\dot{\gamma}} \;+\; \frac{\eta}{\delta \varepsilon_{T_1}} ,
\end{equation}
which decreases with increasing shear rate $\dot{\gamma}$. The model therefore predicts an apparent shear-thinning behavior, specifically due to the elastic offset.

We then explain the shear-thinning behavior observed on $\tau$ combining Eq. (\ref{eq:apperentshearthinning}) and an analytical model that links the emergent shear strain $\dot{\gamma}$ with the active parameters $\mu_a$ and $v_a$. Adapting the polar active vertex model which led to Eq. (\ref{eq:vfinal}) to double-rings, we find that the final rotation velocity in each ring reads
\begin{equation} \label{eq:v2}
V_2 = v_a \frac{\mu_a B  r}{\sqrt{\mu_a^2 r^2 B^2 +v_a^2}} , 
\end{equation}
where $B \in (0,1)$ is a geometric constant that depends on the ring overlap thickness $\ell$ and flow structure (see Supplemental Material \cite{SM}); $B$ being larger in the ODR than SDR setup, we predict that flows are faster in the ODR case than in the SDR case, as observed in experiments, see Fig. \ref{fig:1}F. The agreement of Eq. (\ref{eq:v2}) with vertex model simulations is quantitative, see SM \cite{SM}. In addition, $V_2$ decreases with $1/\mu_a$ and increases with $v_a$; with $\dot{\gamma}\propto V_2$ and $\tau \propto \eta_{\mathrm{app}}$, as defined in Eq. (\ref{eq:apperentshearthinning}), this explains the trend observed of $\tau$ increasing with $1/\mu_a$ and decreasing with $v_a$ in simulations, Fig. \ref{fig:5}D-E. 

\newparagraph{Discussion: biological implications}

\newsubparagraph{WT parameter values}
Our active polar force model yields an estimate of the alignment strength $\mu_{a}$ in the WT case that is comparable to the value $\mu_a= 0.06$ h$^{-1}$ proposed in Ref. \cite{lin2018dynamic} to model WT MDCK.

\newsubparagraph{NMII-deficient cell activity and deformability} 
Our data indicate that depletion of NMIIA or NMIIB reduces migration speed. For NMIIA, this contradicts the trend seen for isolated MDCK cells within unconfined 2D environments \cite{Doyle2012,Jorrisch2013}, but agrees with experiments in 1D channel confinement \cite{Doyle2012}. The ring geometries used here are indeed close to the 1D channel geometries used in \cite{Doyle2012}. Upon considering unconfined 2D migration and migrating isolated cells, Ref. \cite{Halder2019} reported a marked increase in directional persistence, which would be consistent with our results, where both NMIIA- and NMIIB-depleted cells tend to exhibit enhanced directionality despite opposite effects on speed. In addition, NMIIB has been shown to be required to maintain a front-back polarity \cite{Jorrisch2013,10.7554/eLife.46599};  this is also consistent with our result that the active traction strength $v_a$ and polarity coupling $\mu_a$ are reduced in ShIIB cells. While in \cite{10.7554/eLife.46599}, silencing of myosin IIB was shown to weaken cell-cell junctions, our result is consistent with \cite{Chan2015}, whereby individual cells with blocked myosin IIB have been proposed to be less deformable. 

\section*{Perspectives}

Migrating epithelial monolayers typically exhibit turbulent-like or transient flows, which obscure the relationship between cell deformation and cell rearrangements. Here, we used an experimental setup that constrained the flow to a relatively steady-state regime. The success of the tissue–kinematics framework in our context suggests that it could be extended to study turbulent regimes -- by considering a co-moving frame attached to the interface between neighboring vortices. Indeed, such interface displays shear flows and cell-shape patterns closely resembling those in the two-ring overlap region studied here.  A more detailed formulation of this co-moving-frame approach will be presented in future work.

In this context, our method could be used to infer mechanical parameters from self-organized flows in a wide variety of active tissues and cell assemblies, as large-scale vortices are common both \textit{in~vitro} — e.g. with multiple vortices emerging in MDCK tissues once domains exceed a few hundred micrometers in diameter~\cite{Ladoux2017,Heinrich2020} — and \textit{in~vivo}, e.g. the \textit{Polonaise} flows of \textit{Drosophila} gastrulation~\cite{Streichan2018} and early avian development~\cite{Saadaoui2020}. In active fluids, interacting vortices have been characterized in confined colloidal rollers~\cite{Jorge2024} and bacterial suspensions~\cite{Wioland2016}, revealing a possible switch from positive (e.g. more ODR than SDR) to negative (e.g. more SDR than ODR) correlations between the signs of neighboring vortices. In such systems, a Voronoi tessellation of particle centers defines deformable “cells” whose rheology could, in principle, be analyzed using the tissue–kinematics framework introduced here. We plan to develop this analogy in future work.

\section*{Conclusions}

Our work introduces the \textit{live rheometer} setup, a two-step protocol that combines cell migration across two distinct micropatterned geometries, enabling the noninvasive mechanical characterization of epithelial monolayers. By comparing wild-type experiments with dynamical vertex-model simulations, we find that cell–cell viscosity accounts for the onset of an emergent Maxwell-type viscoelastic behavior. We further apply our protocol to show that perturbations of myosin~II isoforms NMIIA and NMIIB affect both the activity and deformability of multicellular assemblies. Finally, we show that migration speed and cell–cell viscosity together determine the tissue viscoelastic relaxation time, i.e., the extent of elastic deformation that cells sustain before undergoing a T1 transition. Such regulation of viscoelasticity could underlie the rheological transition observed during development, from a dynamic, fluid-like state to a more rigid, solid-like architecture~\cite{Mongera2018}.


We expect our noninvasive method for inferring tissue rheology to be broadly applicable. First, the two-step protocol we introduce is versatile and can be adapted to investigate other tissues or factors influencing collective cell migration. Second, by leveraging self-generated forces at sheared interfaces—for instance, between vortex domains—this framework lays the groundwork for a noninvasive characterization of fluidity in the broader context of active granular materials.

\vskip1cm

\small{
\section*{Methods}

\methodparagraph{Cell culture and reagents}
MDCK strain II cells were cultured in DMEM medium plus 10\%  fetal bovine serum and 1\% penicillin–streptomycin at 37 degree celsius with \%5 CO2. Cells were subjected to Mitomycin-C to block  cell division (treatment for $1$h at 10 $\mu$g.ml$^{-1}$ at the beginning of the experiment). MDCK strains silenced for Myosin IIA (ShIIA) and Myosin IIB (ShIIB) have previously been described in Ref. \cite{10.7554/eLife.46599}.

\methodparagraph{Micropattern preparation for collective cell migration on line-patterned strips} A microcontact printing method on polydimethylsiloxane (PDMS) was used to generate the fibronectin-coated pattern confining cell migration \cite{Piel2014, Vedula2014,Jain2020}.

\methodparagraph{Imaging: phase contrast}
Live cell dynamics was observed with phase contrast at 10x and 20x magnification using a Nikon Biostation IMQ microscope. The time-lapse was performed at $37^\circ$ and 5\% CO2. The time-lapse was usually done for $10 \, \mathrm{min}$ intervals and up to $48-50 \, \mathrm{h}$ of acquisition.

\methodparagraph{Data display and statistics} If not stated otherwise, we use the two-tailed unpaired t-test to estimate the P values.
Prism (GraphPad Software) and Matlab (Math Works) were used for data analysis and graph plotting. ANOVA test and Student's t-test paired or unpaired were carried out to analyze the levels of significant differences.

\methodparagraph{Image analysis} For the shear rate estimation, if not stated otherwise, we use the optical flow method previously presented in Ref. \cite{Tlili2020}. For the cell shape analysis, we used Cellpose TissueNet \cite{stringer2021cellpose,Pachitariu2022} for the estimation of the cell strain (similar results were obtained with the most recent Cellpose version, \cite{Pachitariu2025}, which we benchmark against a manually annotated datasets). Details in the Supplemental Material \cite{SM}.

\methodparagraph{Simulations implementation}
We initialize our simulations by placing cell centers along a regular hexagonal pattern along the target (single or double-ring) pattern. Each cell is assigned a target area whose value is set according to a Gaussian noise (with 5$\%$ standard deviation), mimicking the observed cell-to-cell variability in cell area observed in experiments. In the double-ring case, to ensure that the tissues from each ring interact, we consider a slight overlap between the rings (see details in the Supplemental Material \cite{SM} and SI Table S1 for the default set of the parameters).

\methodparagraph{Individual cell strain tensor} Following \cite{Graner2008}, the cell strain is defined at the individual cell level as
\begin{equation}
\bm{\varepsilon}_{J} = \frac{1}{2} \frac{\bm{M}_J-\bm{M}_0}{\bm{M}_0}, \label{eq:varepsilonJ}
\end{equation} 
where $\bm{M}_J$ is the texture tensor of the $J$-th cell, defined as 
\begin{equation}
\bm{M}_J = \frac{1}{N_J} \sum_{K} \left(\bm{r}_K - \bm{r}_J \right) \otimes \left(\bm{r}_K - \bm{r}_J \right) ,     
\end{equation}
with the index $K$ spanning over the $N_J$ neighbor cells to the cell $J$; $\bm{M}_0$ represents a reference texture tensor \cite{Graner2008}, which we set to be one of the regular hexagonal cell patterns in a stress-free state. Averaging cell segmentation masks, we set $\bm{M}_0 = (32.5\mu$m)$^2$ for all experiments.

\methodparagraph{Texture field $\bm{M}$}
We define the strain tensor $\bm{\varepsilon}_{\rm cell}$ through the following averaging procedure \cite{Graner2008}:
\begin{equation} \label{def:Mmatrix}
\bm{\varepsilon}_{\rm cell}\left( \bm{r} \right)= \left\langle \bm{\varepsilon}_J \right\rangle = \frac{\sum\limits_{\left| \bm{r}-{{\bm{r}}_{J}} \right|<{{r}_{\text{cut-off}}}}{w\left( \bm{r}-{{\bm{r}}_{J}} \right)\bm{\varepsilon}_J}}{\sum\limits_{\left| \bm{r}-{{\bm{r}}_{J}} \right|<{{r}_{\text{cut-off}}}}{w\left( \bm{r}-{{\bm{r}}_{J}} \right)}} , 
\end{equation}
where $\bm{\varepsilon}_J$ is defined in Eq. (\ref{eq:varepsilonJ}) and $w(\bm{r}-\bm{r}_J)$ is a weight function, defined as a truncated Gaussian function, 
\begin{equation}
w\left( \bm{r}-{{\bm{r}}_{J}} \right)=\dfrac{1}{\sqrt{2\text{ }\!\!\pi\!\!\text{ }}\sigma }\exp \left( -\dfrac{1}{2}\dfrac{{{\left| \bm{r}-{{\bm{r}}_{J}} \right|}^{2}}}{{{\sigma }^{2}}} \right). \label{eq:WeightFunction}
\end{equation}
and ${{\bm{r}}_{J}} = \sum_{j\in cell}{{{\bm{r}}_{j}}}/n_J$ is the geometric center of the vertices of the cell $J$.
In both experiments and simulations, we set the kernel size at $\sigma = 0.75$ and the cut-off length at $r_{\rm cut-off} = 3\sigma = 2.25$ in the units of a characteristic cell length \cite{SM}.

\methodparagraph{Cell shape analysis} The cell elongation metric is based on the cell area tensor. The cell area tensor is based on a linear weight density along the segments of the cell boundaries, i.e., whose first component reads $I_{xx} = \iint (x - \bar{x})^2 \, dx \, dy$,  where $(\bar{x}, \bar{y})$ is the position of the barycenter of the cell, and $(x,y)$ spans over the cell boundaries. Following Eq. (\ref{def:Mmatrix}), we average the tensor $I$ over the kernel $w$ defined in Eq. (\ref{eq:WeightFunction}). The quantity called cell elongation is then estimated as
\begin{equation}
\varepsilon_{\mathrm{el}} = \log\big(\sqrt{\lambda_1/\lambda_2}\big),
\end{equation}
where $\lambda_1$ (resp. $\lambda_2$) is the maximum (resp. minimum) eigenvalue of the average tensor $I$.

\methodparagraph{Optic flow} The single-ring velocities and indirect cell rearrangement rate tensor inference (following \cite{Tlili2020}, see SM \cite{SM}) are obtained using the optic flow algorithm with KLT algorithm \cite{Tlili2020}. We mask irrelevant regions and filter out all the velocities below a threshold. 

\methodparagraph{Cell rearrangement rate tensor} We track every segmented cell using the btrack algorithm \cite{ulicna2021automated}. We correct those tracking with manual annotation on our Napari pipeline \cite{sofroniew2022}. We compute changes in connectivity among masks to identify instances of cell-cell junction disappearance or appearance. Following Ref. \cite{Graner2008}, we estimate the cell rearrangement rate tensor at the location $\bm{r}$ using the expression:
\begin{equation} \label{eq:Tfield}
\bm{T}(\bm{r}) = \frac{N_{\mathrm{T1}}}{\Delta t N_{\mathrm{link}}} 
\sum\limits_{j=1, \ldots, N_{\mathrm{T1}}}{w(\bm{r}-{{\bm{r}}_{j}}) (\bm{\ell}^{j}_{a} \otimes \bm{\ell}^{j}_{a}-\bm{\ell}^{j}_{d} \otimes \bm{\ell}^{j}_{d})} , 
\end{equation}
where $\Delta t$ is the observation duration; $\bm{\ell}_a$ and $\bm{\ell}_d$ are vectors between cell centers of appearing or disappearing cell--cell edges, respectively;
$N_{\mathrm{T1}}$ is the total number of rearrangements that occurred; $N_{\mathrm{link}}$ is the number of cell-cell links; both $N_{\mathrm{T1}}$ and $N_{\mathrm{link}}$ are weighted-average around $\bm{r}$, with $w$ the averaging kernel defined in Eq. (\ref{eq:WeightFunction}). 
From the coarse-grained tensors $\bm{M}$ and $\bm{T}$, we quantify the cell rearrangement rate tensor field based on T1 topological transition events as
\begin{equation}
\dot{\bm{\varepsilon}}_{r} = -\frac{1}{2} \big( \bm{M}^{-1}\cdot\bm{T} \big)_{\mathrm{sym}} , \label{eq:CellRearrangementRateTensor}
\end{equation}
where $()_{\mathrm{sym}}$ is the tensor symmetrization operator \cite{Graner2008}.
}

\section*{Acknowledgement}{JFR thanks Laura Dalla Pozza for improving the one-ring analysis. We would like to thank all the members of the "Cell adhesion and Mechanics” team for helpful discussions, as well as A. Kabla, H. Delanoe-Ayari and F. Graner. This work was supported by the European Research Council (Grant No. Adv-101019835 to BL), LABEX Who Am I? (ANR-11- LABX-0071 to BL and RMM), the Ligue Contre le Cancer (Equipe labellisée 2019 to RMM), the Alexander von Humboldt Foundation (Alexander von Humboldt Professorship to BL), Institut National du Cancer (INCa 18429 to BL and RMM), the Agence Nationale de la Recherche (“STRATEPI” DFG-ANR-22-CE92-0048) to RMM). We acknowledge the ImagoSeine core facility of the IJM, a member of IBiSA and France-BioImaging (ANR-10-INBS-04)
infrastructures. The project leading to this publication has received funding from France 2030, the French Government program managed by the French National Research Agency (ANR-16-CONV-0001) from Excellence Initiative of Aix-Marseille University - A*MIDEX. R-M.M. is also funded by ANR-17-CE13-0013 and J.-F. R. by ANR-20-CE30-0023. G.H.N. and S.P. were recipients of the Labex Who Am I?}.

\bibliographystyle{apsrev4-2}
\bibliography{tworings}

\end{document}